\documentclass[conference]{IEEEtran}

\usepackage{amsmath}
\usepackage{amssymb}
\usepackage{graphicx}
\usepackage{latexsym}
\usepackage{setspace}
\usepackage{mathtools}
\usepackage{amsmath}
\usepackage{yhmath}
\usepackage{graphicx,dblfloatfix}
\usepackage{cite}
\usepackage[caption=false,labelformat=simple,font=footnotesize,position=bottom]{subfig}
\usepackage{setspace}
\usepackage{bm}
\usepackage{upgreek}
\usepackage{multirow}
\usepackage{afterpage}
\usepackage{makecell}
\usepackage[T1]{fontenc}
\usepackage{booktabs}
\usepackage{threeparttable}
\usepackage[table]{xcolor}
\usepackage{arydshln}
\usepackage{bbding}
\usepackage{amssymb}
\usepackage{amsbsy}
\usepackage{tikz}
\usepackage[hidelinks]{hyperref}
\usepackage{eso-pic} 
\def\BibTeX{{\rm B\kern-.05em{\sc i\kern-.025em b}\kern-.08em
    T\kern-.1667em\lower.7ex\hbox{E}\kern-.125emX}}

\usetikzlibrary{arrows.meta,backgrounds,calc,fit,positioning,shapes.geometric,trees}
\usepackage{algpseudocode}

\definecolor{myred}{rgb}{0.7686,0.0588,0.0941}
\definecolor{mygreen}{rgb}{0.2000,0.5804,0.1804}
\definecolor{myyellow}{rgb}{0.9,0.8,0}
\definecolor{myyellow2}{rgb}{0.8,0.6,0}
\definecolor{mygray}{rgb}{0.92,0.92,0.92}
\definecolor{mygray2}{rgb}{0.8,0.8,0.8}
\definecolor{mymagenta}{rgb}{0.9999,0.5,0.9999}
\definecolor{mymagenta2}{rgb}{0.7,0.2,0.7}
\definecolor{myGreen}{rgb}{0,0.9,0.7}
\definecolor{myGreen2}{rgb}{0,0.65,0.45}
\definecolor{mygreen3}{rgb}{0.76,0.83,0.61}
\definecolor{myblue}{rgb}{0.6,0.6,1}
\definecolor{myblue2}{rgb}{0.3,0.3,0.9}
\definecolor{myblue3}{rgb}{0.73,0.8,0.90}
\definecolor{myorange}{rgb}{0.9804,0.7137,0.4941}
\definecolor{myorange2}{rgb}{0.9922,0.9020,0.8275}
\definecolor{myorange3}{rgb}{0.99, 0.84,0.71}
\definecolor{bluebl}{HTML}{DBDBDB}
\definecolor{myDarkPurple}{RGB}{120,0,120}
\definecolor{flowgray}{RGB}{229,229,229}
\definecolor{flowred}{RGB}{247,151,154}
\definecolor{flowblue}{RGB}{170,203,231}
\definecolor{flowdarkred}{RGB}{145,52,58}
\definecolor{flowdarkblue}{RGB}{45,83,125}
\definecolor{flowgreen}{RGB}{196,219,162}
\definecolor{flowdarkgreen}{RGB}{70,108,48}
\definecolor{flowyellow}{RGB}{255,229,168}
\definecolor{floworange}{RGB}{248,188,145}
\definecolor{flowpurple}{RGB}{218,199,247}

 \newcommand{\Unit}[1]{\ensuremath{{\:}\mathrm{#1}}}

\newcommand{\A}{\Unit{A}}

\newcommand{\V}{\Unit{V}}

\newcommand{\mH}{\Unit{mH}}

\newcommand{\ohm}{\Unit{{\Omega}}}

\newcommand{\Nm}{\Unit{Nm}}
\newcommand{\rpm}{\Unit{r/min}}

\newcommand{\sumsmallb}{\mathrel{\scalebox{0.5}{$+$}}}
\newcommand{\minsmallb}{\mathrel{\scalebox{0.5}{$-$}}}

\newcommand{\eqsupref}[1]{\mathrm{(\ref{#1})}}

\makeatletter
\newif\ifnobrackets
\renewcommand\@cite[2]{\ifnobrackets\else[\fi{#1\if@tempswa , #2\fi}\ifnobrackets\else]\fi\nobracketsfalse}

\makeatother

\makeatletter
\newcommand{\raisemath}[1]{\mathpalette{\raisem@th{#1}}}
\newcommand{\raisem@th}[3]{\raisebox{#1}{$#2#3$}}
\makeatother

\makeatletter
\renewcommand{\Function}[2]{%
  \csname ALG@cmd@\ALG@L @Function\endcsname{#1}{#2}%
  \def\jayden@currentfunction{#1}%
}
\newcommand{\funclabel}[1]{%
  \@bsphack
  \protected@write\@auxout{}{%
    \string\newlabel{#1}{{\jayden@currentfunction}{\thepage}}%
  }%
  \@esphack
}
\makeatother

\graphicspath{{Figs/}}

\makeatletter
\def\blfootnote{\xdef\@thefnmark{}\@footnotetext}
\makeatother

\makeatletter
\def\ignore#1{%
     \begingroup
         \@fileswfalse
         #1
    \endgroup
}
\makeatother

\makeatletter
\newcommand{\linebreakand}{%
  \end{@IEEEauthorhalign}
  \hfill\mbox{}\par
  \mbox{}\hfill\begin{@IEEEauthorhalign}
}
\makeatother

\makeatother

\begin{document}

\title{Current References for Minimum Copper Loss and Torque Ripple in the Full Torque-Speed Range for Symmetrical Six-Phase PMSMs With Nonsinusoidal Back-EMF Under an Open-Phase Fault\\
\thanks{This work was supported in part by the Spanish State Research Agency (AEI) under project PID2022-136908OBI00/MCIN/AEI/10.13039/501100011033/FEDER-UE.}}

\author{
\IEEEauthorblockN{Alejandro G. Yepes\textsuperscript{1}, Wessam E. Abdel-Azim\textsuperscript{1,2}, Oscar López\textsuperscript{1}, \\ Petros Karamanakos\textsuperscript{3}, Ayman S. Abdel-Khalik\textsuperscript{4}, and Jesús Doval-Gandoy\textsuperscript{1}}\vspace{4pt}

\IEEEauthorblockA{\textsuperscript{1}\textit{CINTECX}, \textit{Universidade de Vigo}, \textit{APET}, Vigo, Spain}

\IEEEauthorblockA{\textsuperscript{2}\textit{Department of Electrical Engineering}, \textit{Alexandria University}, Alexandria, Egypt}

\IEEEauthorblockA{\textsuperscript{3}\textit{Faculty of Information Technology and Communication Sciences}, \textit{Tampere University}, Tampere, Finland} 

\IEEEauthorblockA{\textsuperscript{4}\textit{Department of Electrical and Computer Engineering}, \textit{Sultan Qaboos University}, Muscat, Oman}

\IEEEauthorblockA{Email: agyepes@uvigo.es, wessam.essam@uvigo.es, olopez@uvigo.es,  \\ p.karamanakos@ieee.org, a.abdelkhalik@squ.edu.om, jdoval@uvigo.es}\vspace{-11pt}
}

\IEEEoverridecommandlockouts

\maketitle

\AddToShipoutPictureFG*{%
  \AtPageUpperLeft{%
    \raisebox{-5mm}[0pt][0pt]{%
      \makebox[\paperwidth][c]{%
        \scriptsize\itshape
        Accepted for presentation at IECON 2026, the 52nd Annual Conference of the IEEE Industrial Electronics Society.%
      }%
    }%
  }%
  \AtPageLowerLeft{%
    \raisebox{5mm}[0pt][0pt]{%
      \makebox[\paperwidth][c]{%
        \parbox[b]{0.94\paperwidth}{\centering\scriptsize
          \textcopyright{} 2026 IEEE. Personal use of this material is permitted.
          Permission from IEEE must be obtained for all other uses, in any current
          or future media, including reprinting/republishing this material for
          advertising or promotional purposes, creating new collective works, for
          resale or redistribution to servers or lists, or reuse of any copyrighted
          component of this work in other works.}%
      }%
    }%
  }%
}%
\thispagestyle{empty}
\pagestyle{empty} 
\begin{abstract}
Current-reference generation based on lookup tables (LUTs) is here proposed for star-connected symmetrical six-phase nonsalient PMSMs with nonsinusoidal back-EMF under an open-phase fault. Fourier coefficients for all healthy phases are computed offline, enabling unbalanced nonsinusoidal currents. A  lexicographic optimization minimizes torque ripple and then copper loss, subject to torque, zero-current-sum, peak-current, peak-voltage, and torque-ripple constraints. Cogging torque can be included for compensation. Simulations show a higher feasible speed limit: about 30\% at low torque and 23\% along an example load curve.  Finite-element analysis shows a 77\% reduction in peak-to-peak torque ripple with cogging-torque compensation.
\end{abstract}


\section*{Nomenclature}
\setlength\tabcolsep{0pt}\vspace{-1pt}

\noindent\begin{tabular}{@{\extracolsep\fill}p{0.12\columnwidth}p{0.88\columnwidth}}
$\bm{0}_{\mu \times \nu}$ & Null matrix of size $\mu$$\times$$\nu$.\\
$\bm{1}_{\mu \times \nu}$ & All-ones matrix of size $\mu$$\times$$\nu$.\\
$\bm{A}$ & Constraint matrix of the optimization problems.\\ 
$\bm{b}$ & Constraint vector of the optimization problems.\\ 
$\bm{D}$ & Six-phase vector space decomposition (VSD).\\ 
$e_{k}(t)$ & Back-electromotive force (back-EMF) in phase $k$. \\
$e^{\prime}_{k}(t)$ & Back-EMF $e_{k}(t)$ normalized by mechanical speed. \\
$h$ & Selected current harmonic orders, within $\left\{1,3,...,H\right\}$. \\
$i_{k}(t)$ & Instantaneous current in stator phase $k$. \\
$i_{\mathrm{pk}}$ & Per-cycle maximum of phase-current peaks. \\
$i^{\mathrm{mx}}_{\mathrm{pk}}$ & Maximum admissible peak current of the drive, excluding switching ripple. \\
$I^{\mathrm{Re,Im}}_{k,h}$ & Fourier coefficients of $h$th phase-$k$ current harmonic. \\
$J_{\mathrm{SCL}}$ & SCL normalized by stator resistance. \\
$k$ & Healthy stator phase (2 for phase~b, 3 for c, etc.). \\
$\bm{G}^{\prime}$ & Selection matrix of line-voltage signals. \\
$n_{\mathrm{h}}$ & Number of current harmonics considered. \\
$n_{\mathrm{t}}$ & Number of time samples in offline optimization. \\
\end{tabular}

\noindent\begin{tabular}{@{\extracolsep\fill}p{0.12\columnwidth}p{0.88\columnwidth}}
$\theta_{\mathrm{r}}(t)$ & Rotor electrical position. \\
$t$ & Time. The argument $(t)$ is included only for variables whose values remain time-varying for fixed $T^{\ast}$ and $\omega$. \\
$T(t)$ & Electromagnetic torque. \\
$T^{\ast}$ & Reference for $T$, prior to computing current references. \\
$\tau$ & Peak-to-peak $T$ ripple.\\
$\tau^{\min}$ & Minimum achievable $\tau$  per operating point $(T^{\ast},\omega)$.\\
$\tau^{\mathrm{mx}}$ & Maximum acceptable $\tau$ value, never to be surpassed.\\
$v_{k}(t)$ & Instantaneous phase voltage in stator phase $k$. \\
$\bm{v}^{\prime}(t)$ & Instantaneous vector of healthy line voltages. \\
$v_{{\mathrm{pk}}}$ & Maximum peak of $\bm{v}^{\prime}(t)$ per electrical cycle. \\
$v^{\mathrm{mx}}_{{\mathrm{pk}}}$ & Maximum admissible $v_{{\mathrm{pk}}}$ (given by dc-link voltage). \\
$\omega$ & Fundamental electrical frequency (speed). \\
$\omega_{\mathrm{m}}$ & Fundamental mechanical frequency (speed). \\
$\bm{\omega}^{\downarrow}$ & For each $T^{\ast}$, lowest $\omega$ for which $\bm{x}$ varies with $\omega$, i.e., lowest $\omega$ for which the $v^{\mathrm{mx}}_{{\mathrm{pk}}}$ constraint is active. \\
$\bm{\omega}^{\uparrow}$ & For each $T^{\ast}$, highest $\omega$ for which a solution exists. \\
$\bm{x}$ & Vector of optimization variables.\\
$\bm{x}_{\mathrm{opt}}$ & Lexicographic-optimization solution for given $(T^{\ast},\omega)$.\\
$z_{1}, z_{2}$ & Auxiliary variables to limit $\tau$ in the optimization.\\
\end{tabular}

\smallskip

Boldface denotes vectors and matrices.

\section{Introduction}
\label{sec:introduction}

High torque density and reliability motivate the use of multiphase permanent-magnet synchronous machines (PMSMs) \cite{Yepes2022Machines1,Taha2023TPEL}. The six-phase option balances these advantages and complexity \cite{Taha2023TPEL}. Torque density can be further increased with nonsinusoidal back-electromotive-force (back-EMF) designs \cite{Wang2024TIE,Cervone2021TPEL}, typically based on fractional-slot concentrated windings with usually small saliency \cite{Fan2020TVT,Chong2010IET,Zhang2018TAS}. Symmetrical windings are preferred over asymmetrical ones for fault tolerance \cite{Munim2017TPEL} and current quality \cite{Prieto2022TIE}, while a single neutral point maximizes postfault torque range compared with two neutrals \cite{Che2014TPEL,Munim2017TPEL}.

\begin{figure}[t!]
\centering
\includegraphics[width=0.9\columnwidth]{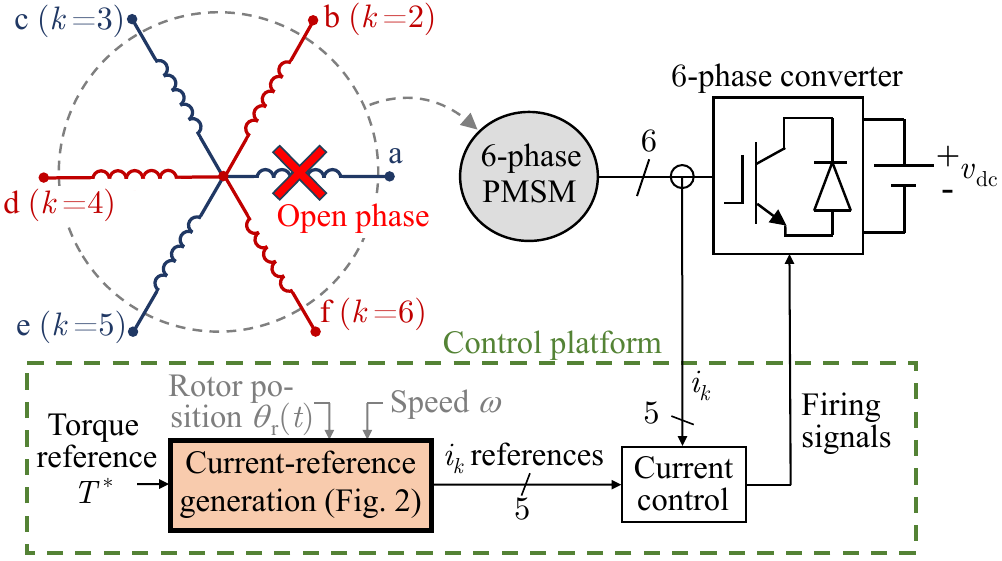}
\caption{Star-connected symmetrical six-phase PMSM drive based on field-oriented control under an open-phase fault in phase~a, including current-reference generation, on which this paper is focused.}
\label{fig:drive}
\end{figure}

For PMSMs driven by field-oriented control (Fig.~\ref{fig:drive}), current references for the inner current loop must be generated from the torque command, rotor speed and position. In \cite{Yepes2024TTE}, an online method, the nonsinusoidal-back-EMF full-torque-range minimum-loss (NSBE-FRML) strategy, was proposed for nonsalient PMSMs with any phase number, arbitrary back-EMF waveform, symmetrical or asymmetrical windings, and healthy or open-phase operation. It provides nonsinusoidal current references minimizing stator copper loss (SCL) and torque ripple under current and torque-ripple limits. However, it is not able to fully exploit the available speed.

Current-reference generation for PMSMs under voltage constraints has been addressed by several works \cite{Lu2026TPEL,Atallah2003TIA,Sun2010TIE,Vu2019IET,Moraes2020EPE,Laksar2026TIE,Lu2026TTE}. As reviewed in \cite{Yepes2026TIE}, these solutions do not meet all targeted requirements. Some cannot accommodate arbitrary back-EMF harmonic content \cite{Lu2026TPEL,Moraes2020EPE,Laksar2026TIE,Lu2026TTE}, do not consider current harmonic injection \cite{Lu2026TPEL,Lu2026TTE}, or are not applicable to symmetrical six-phase PMSMs \cite{Lu2026TPEL,Vu2019IET,Moraes2020EPE,Laksar2026TIE,Lu2026TTE}. Others use fewer control degrees of freedom, reducing efficiency and torque-speed capability \cite{Atallah2003TIA,Sun2010TIE}, or do not exploit an admissible torque-ripple margin to enlarge torque-speed region \cite{Lu2026TPEL,Atallah2003TIA,Sun2010TIE,Vu2019IET,Moraes2020EPE,Laksar2026TIE,Lu2026TTE}.


To address these limitations, \cite{Yepes2026TIE} introduced an alternative current-reference generation strategy for nonsalient PMSMs. It retains the generality of \cite{Yepes2024TTE} regarding phase number, back-EMF waveform and winding arrangement, while incorporating the voltage limit to minimize SCL and torque ripple over the full torque-speed region \cite{Yepes2026TIE}, thereby overcoming the limitations of \cite{Lu2026TPEL,Atallah2003TIA,Sun2010TIE,Vu2019IET,Moraes2020EPE,Laksar2026TIE,Lu2026TTE}. Optimal Fourier coefficients of the current references are obtained offline by quadratic optimization and stored in lookup tables (LUTs); online, the LUTs generate the references from torque command and speed.


However, the method from \cite{Yepes2026TIE} has several drawbacks. Open-phase faults, such as those caused by converter-leg, line or winding open-circuit faults, or by remedial actions to isolate a faulty part of one phase or leg \cite{Yepes2022Machines2,Wang2024TIE,Munim2017TPEL,Lu2026TPEL}, were not considered. Moreover, despite substantially simplifying the offline optimization by assuming healthy balanced conditions, the LUT generation process took multiple hours. This was because the optimization problem was repeatedly solved while increasing the allowed torque ripple  from nearly zero in very small steps until feasibility was reached. In addition, the torque ripple due to cogging torque was not compensated.

This paper addresses these gaps by improving the current-reference generation method from \cite{Yepes2026TIE}, for the case of a six-phase PMSM with symmetrical star-connected windings and a single neutral point. An open-phase fault is considered (see Fig.~\ref{fig:drive}), so that minimum SCL, smooth torque, maximum (full) torque-speed range, and compliance with the drive constraints are fulfilled despite the postfault imbalance. The additional LUT-generation time caused by the increased number of degrees of freedom is mitigated by reformulating the optimization as a two-stage lexicographic problem for each operating point: a linear program to minimize the torque ripple, followed by a quadratic program to minimize the SCL while limiting to such torque-ripple value. Furthermore, the cogging torque is included in the offline optimization for LUT generation to enable its compensation during online operation. Simulation results confirm the functionality of the proposal.

\section{Model of the Six-Phase PMSM}
\label{sec:model}

The stator voltage equation of a multiphase nonsalient PMSM can be written as \cite{Yepes2026TIE}
\begin{equation}
\label{ec:model}
\bm{D}\bm{v}(t)
=
\bm{R}\bm{D}\bm{i}(t)
+
\bm{L}\tfrac{\mathrm{d}}{\mathrm{d}t}\bm{D}\bm{i}(t)
+
\bm{D}\bm{e}(t)
\end{equation}
where the phase-domain vectors $\bm{v}$, $\bm{i}$ and $\bm{e}$
collect the six stator voltages, currents and back-EMFs, with the general form
\begin{equation}
\bm{u}(t)=
\left[
\begin{array}{cccccc}
u_{1}(t) & u_{2}(t) & \cdots & u_{k}(t) & \cdots & u_{6}(t)
\end{array}
\right]^{\mathrm{T}}
\in\mathbb{R}^{6\times1}.
\end{equation}
The indices $k=1,\ldots,6$ correspond to phases $a$--$f$, respectively. $\bm{D}\in\mathbb{R}^{6\times 6}$ in (\ref{ec:model}) is the vector-space-decomposition (VSD) matrix, which for a six-phase PMSM with symmetrical windings (three-phase sets displaced by $\delta=60^{\circ}$) is \cite{Yepes2025TIE}
\begin{equation}\renewcommand{\arraycolsep}{2pt}
\label{ec:VSD}\bm{D}=\frac{1}{3}\left[\begin{array}{@{}cccccc@{}} 1 & \cos(\delta) & \cos(2\delta) & \cos(3\delta) & \cos(4\delta) & \cos(5\delta) \\ 
0 & \sin(\delta) & \sin(2\delta) & \sin(3\delta) & \sin(4\delta) & \sin(5\delta)\\ 
1 & \cos(2\delta) & \cos(4\delta) & \cos(6\delta) & \cos(8\delta) & \cos(10\delta) \\ 
0 & \sin(2\delta) & \sin(4\delta) & \sin(6\delta) & \sin(8\delta) & \sin(10\delta)\\
1/2 & 1/2 & 1/2 & 1/2 & 1/2 & 1/2 \\ 
1/2 & -1/2 & 1/2 & -1/2 & 1/2 & -1/2 \end{array}\right]\vspace{-5pt}
\end{equation}
such that\vspace{-8pt}
\begin{equation}\renewcommand{\arraycolsep}{4pt}
\label{ec:mult_by_Tn}
\rule{0pt}{20pt}\left[\begin{array}{cccccc} 
u_{\alpha}(t)  & u_{\beta}(t) & u_{x}(t) & u_{y}(t) & u_{0^{\sumsmallb}}(t) & u_{0^{\minsmallb}}(t) \end{array}\right]^{\mathrm{T}}=\bm{D}\bm{u}(t).
\end{equation}
$\bm{R},\bm{L}\in\mathbb{R}^{6\times6}$ are diagonal matrices of the VSD-axis stator resistances and self-inductances.

The instantaneous electromagnetic torque is \cite{Atallah2003TIA,Kestelyn2011TIE,Mohammadpour2015TIA}
\begin{equation}
\label{ec:Tem}
T(t)=e^{\prime}_{1}(t)i_{1}(t)+e^{\prime}_{2}(t)i_{2}(t)+\ldots+e^{\prime}_{6}(t)i_{6}(t)+T_{\mathrm{cog}}(t)\vspace{-2pt}
\end{equation}
where $T_{\mathrm{cog}}(t)$ is the cogging torque, and $e^{\prime}_{k}(t)=e_{k}(t)/\omega_{\mathrm{m}}$ is the
back-EMF of phase $k$ per unit of mechanical speed $\omega_{\mathrm{m}}$.  

For a phase-$k$ open-phase fault, these model equations can be applied while setting $i_{k}(t)=0$ \cite{Che2014TPEL,Yepes2022Machines2}. 

\section{New Current-Reference Generation Method}
\label{sec:proposed}

The proposed method comprises online and offline stages (Sections~\ref{sec:online} and~\ref{sec:offline}). In the online stage, the current references are
obtained from the rotor position, speed and LUT-stored Fourier coefficients of the
phase-current harmonics. The LUTs are indexed by the operating point
$(T^{\ast},\omega)$, with torque reference $T^{\ast}$ and speed $\omega$, and
are generated offline by solving a lexicographic optimization problem over a grid of
$(T^{\ast},\omega)$ values for the phase-a open-phase-fault case.\footnote{Current references for the cases of open-phase faults in other phases may be obtained by simply shifting them circularly, due to symmetry \cite{Munim2017TPEL,Che2014TPEL}.} 

\subsection{Online Generation of Current References}
\label{sec:online}

\begin{figure}[t!]
\centering
\includegraphics[width=0.75\columnwidth]{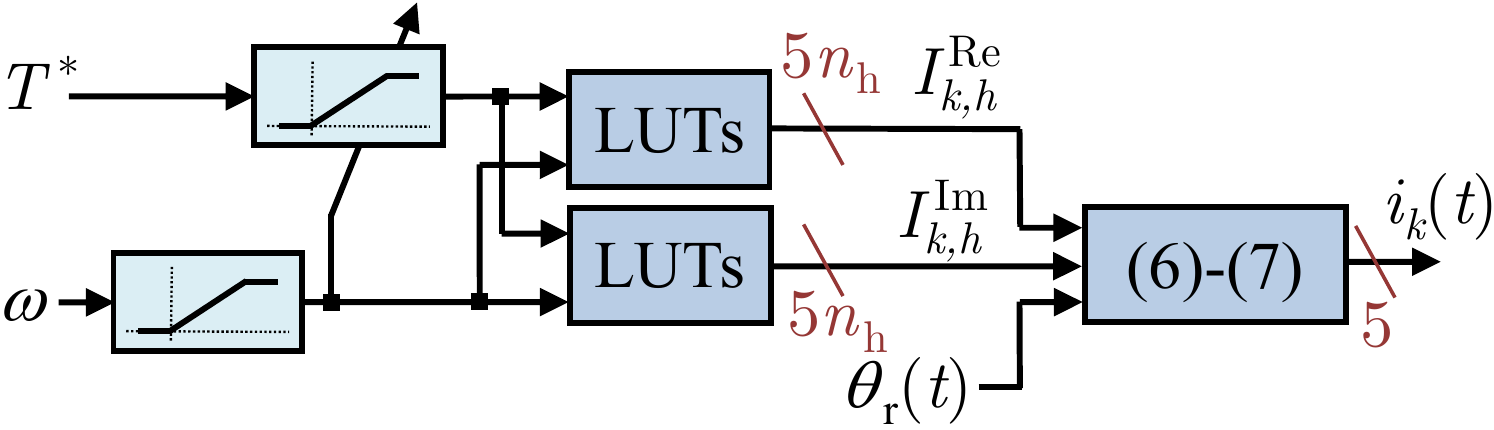}
\caption{Online implementation of the current-reference generation method.}
\label{fig:online_scheme_healthy}
\end{figure}

Fig.~\ref{fig:online_scheme_healthy} shows the online implementation, applied after a phase-a open-phase fault is detected.  Unlike \cite{Yepes2026TIE}, unbalanced currents are enabled by assigning each healthy phase $k\in\{2,3,4,5,6\}$ its own Fourier coefficients $I^{\mathrm{Re}}_{k,h}$ and $I^{\mathrm{Im}}_{k,h}$ for odd harmonics $h\in\{1,3,...,H\}$, while phase~a is open:
\begin{equation}
\label{ec:online}
i_{k}(t)=\sum_{h}I^{\mathrm{Re}}_{k,h}\underbrace{\cos\left(h\theta_{k}(t)\right)}_{c_{k,h}(t)}+
\sum_{h}I^{\mathrm{Im}}_{k,h}\underbrace{\sin\left(h\theta_{k}(t)\right)}_{s_{k,h}(t)}
\end{equation} 
for $k$, whereas $i_1(t)=0$. In (\ref{ec:online}),
\begin{equation}\label{ec:theta_k}
\theta_{k}(t)=\underbrace{\omega t}_{\theta_{\mathrm{r}}(t)}-\underbrace{\left(k-1\right)\frac{\pi}{3}}_{\phi_{k}}.
\end{equation}
The rotor electrical position is $\theta_{\mathrm{r}}(t)$, $\omega$ is the fundamental electrical frequency (assumed constant between control samples), and $\phi_k$ is the electrical phase displacement. The argument $t$ is shown only for quantities that remain time-varying at constant $T^{\ast}$ and $\omega$. In Fig.~\ref{fig:online_scheme_healthy}, $n_{\mathrm{h}}$ denotes the number of current harmonics. For each $(T^{\ast},\omega)$, the LUTs provide the healthy-phase Fourier coefficients $I^{\mathrm{Re}}_{k,h}$ and $I^{\mathrm{Im}}_{k,h}$ in (\ref{ec:online}) by linear interpolation. The LUT inputs $T^{\ast}$ and $\omega$ are saturated to remain inside the tabulated feasible region for the phase-a open-phase-fault case. The LUTs only generate current references (see Fig.~\ref{fig:drive}); hence, closed-loop stability, reference tracking, and disturbance rejection depend on the current controller rather than on the LUTs.

\subsection{Offline LUT Generation}
\label{sec:offline}

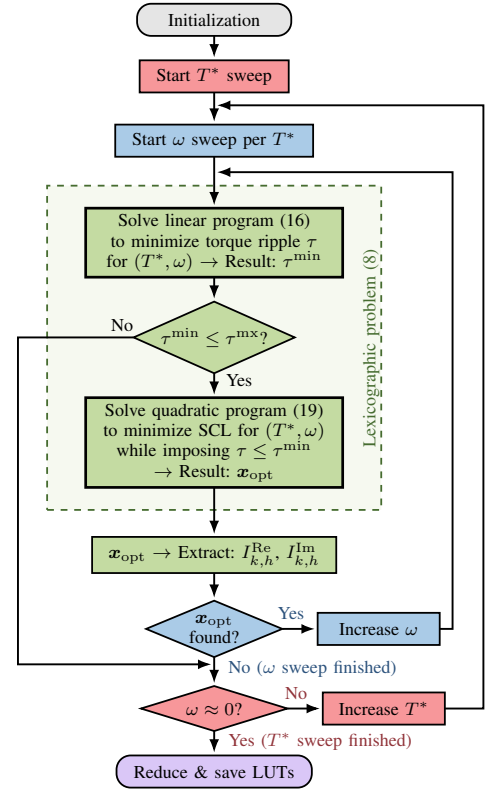
\begin{figure}[t!]
\centering
\scalebox{0.85}{%
\begin{tikzpicture}[
    >=Latex,
    every node/.style={font=\footnotesize, align=center},
    flowline/.style={line width=0.85pt},
    flowarrow/.style={flowline,shorten >=1.2pt,-{Latex[length=2.1mm,width=1.55mm]}},
    flowbox/.style={
        draw=black,
        line width=0.8pt,
        inner xsep=0.07cm,
        inner ysep=0.09cm,
        outer sep=0pt
    },
    terminator/.style={
        flowbox,
        rounded corners=0.22cm,
        minimum height=0.50cm
    },
    process/.style={
        flowbox,
        minimum height=0.5cm
    },
    decision/.style={
        flowbox,
        diamond,
        aspect=2.25,
        inner sep=0.03cm,
        minimum width=2.5cm,
        minimum height=0.7cm
    },
    groupbox/.style={
        draw=flowdarkgreen,
        fill=flowgreen!16,
        dashed,
        line width=0.8pt,
        inner xsep=0.6cm,
        inner ysep=0.35cm
    }
]
    \node[terminator, fill=flowgray, minimum width=2.40cm] (init) {Initialization};
    \node[process, fill=flowred, minimum width=2.32cm, below=0.38cm of init] (startt) {Start $T^{\ast}$ sweep};
    \node[process, fill=flowblue, minimum width=3.08cm, below=0.50cm of startt] (startw) {Start $\omega$ sweep per $T^{\ast}$};

    \node[
        process,
        fill=flowgreen,
				line width=1.2pt,
				minimum width=4cm
    ] (solve_LP) [below=0.8cm of startw] {%
        Solve linear program (\ref{ec:LP})\\ to minimize torque ripple $\tau$ \\
				 for $(T^{\ast},\omega)$ $\rightarrow$ Result: $\tau^{\mathrm{min}}$
    };
		\node[decision, fill=flowgreen, below=0.38cm of solve_LP] (tau_found)
        {\raisebox{0.03cm}{\shortstack[c]{$\tau^{\mathrm{min}}\leq\tau^{\mathrm{mx}}$?}}};
    \node[process, fill=flowgreen, minimum width=4cm, line width=1.2pt, below=0.38cm of tau_found] (solve_QP) {Solve quadratic program (\ref{ec:QP})\\ to minimize SCL for $(T^{\ast},\omega)$\\
		while imposing $\tau\leq\tau^{\mathrm{min}}$\\ $\rightarrow$ Result: $\bm{x}_{\mathrm{opt}}$};
		\node[process, fill=flowgreen, minimum width=3.8cm, below=0.77cm of solve_QP] (extract) {$\bm{x}_{\mathrm{opt}}$ $\rightarrow$ Extract: $I^{\mathrm{Re}}_{k,h}$, $I^{\mathrm{Im}}_{k,h}$};
    \node[decision, fill=flowblue, minimum width=2.12cm, below=0.43cm of extract] (endw) {$\bm{x}_{\mathrm{opt}}$\\ found?};
    \node[decision, fill=flowred, minimum width=2.28cm, minimum height=0.70cm, below=0.45cm of endw] (done) {$\omega\approx0$?};
    \node[process, fill=flowblue, minimum width=1.9cm, right=0.55cm of endw] (incw) {Increase $\omega$};
		\node[process, fill=flowred, minimum width=1.9cm, right=0.55cm of done] (inct) {Increase $T^{\ast}$};

    \node[terminator, fill=flowpurple, minimum width=3.04cm, below=0.38cm of done] (end) {Reduce \& save LUTs};

    \begin{scope}[on background layer]
        \node[groupbox, fit=(solve_LP)(tau_found)(solve_QP)] (lexproblem) {};
    \end{scope}
    \node[font=\footnotesize, text=flowdarkgreen, anchor=south, rotate=90] at ($(lexproblem.east)+(0.08cm,0)$)
        {Lexicographic problem (\ref{ec:lex_compact})};

    \coordinate (loopjoin) at ($(startt.south)!0.40!(startw.north)$);
    \coordinate (loopright) at ($(inct.east)+(0.6cm,0)$);
		\coordinate (loopjoin2) at ($(startw.south)!0.30!(solve_LP.north)$);
		\coordinate (loopright2) at ($(incw.east)+(0.20cm,0)$);
		\coordinate (loopleft) at ($(tau_found.west)-(1.8cm,0)$);
		\coordinate (xopt_Tend) at ($(endw.south)!0.25!(done.north)$);

    \draw[flowarrow] (init.south) -- (startt.north);
    \draw[flowarrow] (startt.south) -- (startw.north);
    \draw[flowarrow] (startw.south) -- (solve_LP.north);
		\draw[flowarrow] (solve_LP.south) -- (tau_found.north);
		\draw[flowarrow] (tau_found.south) -- (solve_QP.north);
		\draw[flowarrow] (solve_QP.south) -- (extract.north);
		\draw[flowarrow] (extract.south) -- (endw.north);
    \draw[flowarrow] (endw.south) -- (done.north);
		 \draw[flowarrow] (endw.east) -- (incw.west);
    \draw[flowarrow] (done.south) -- (end.north);
    \draw[flowarrow] (done.east) -- (inct.west);
    \draw[flowarrow] (inct.east) -- (loopright) -- (loopright |- loopjoin) -- (loopjoin);
		\draw[flowarrow] (incw.east) -- (loopright2) -- (loopright2 |- loopjoin2) -- (loopjoin2);
		\draw[flowarrow] (tau_found.west) -- (loopleft) -- (loopleft |- xopt_Tend) -- (xopt_Tend);

		\node[font=\footnotesize, anchor=west] at ($(tau_found.south)!0.3!(solve_QP.north)+(0.06,0)$) {Yes};
		\node[font=\footnotesize, anchor=south] at ($(tau_found.west)-(0.18,0.02)$) {No};
		\node[font=\footnotesize, text=flowdarkblue, anchor=west] at ($(endw.south)!0.42!(done.north)+(0.1,0)$) {No ($\omega$ sweep finished)};
    \node[font=\footnotesize, text=flowdarkblue, anchor=south] at ($(endw.east)!0.25!(incw.west)+(0,0.02)$) {Yes};
		\node[font=\footnotesize, text=flowdarkred, anchor=west] at ($(done.south)!0.42!(end.north)+(0.1,0)$) {Yes ($T^{\ast}$ sweep finished)};
		\node[font=\footnotesize, text=flowdarkred, anchor=south] at ($(done.east)!0.25!(inct.west)+(0,0.02)$) {No};
\end{tikzpicture}%
}
\caption{Flowchart of the offline stage of the proposed current-reference generation method, to generate the LUTs for the phase-a-open case. The colors indicate initialization (gray), $T^{\ast}$ sweeps (red), $\omega$ sweeps per $T^{\ast}$ (blue), lexicographic optimization problem (green), and finalizing the LUTs (purple).}
\label{fig:offline_flowchart}
\end{figure}

Fig.~\ref{fig:offline_flowchart} depicts the offline LUT-generation process. For each operating point $(T^{\ast},\omega)$, a lexicographic optimization problem finds the healthy-phase coefficients $I^{\mathrm{Re}}_{k,h}$ and $I^{\mathrm{Im}}_{k,h}$, to be stored in the LUTs per $(T^{\ast},\omega)$. A compact statement of this lexicographic problem is first given in Section~\ref{sec:compact_lex}. This lexicographic problem is then decomposed into two optimization stages, detailed in Sections~\ref{sec:LP_stage} and~\ref{sec:QP_stage}. The overall offline process to generate the LUTs for all operating points is explained in Section~\ref{sec:LUT}. 

\subsubsection{Compact Statement of Lexicographic Optimization}
\label{sec:compact_lex}

A lexicographic problem to find the healthy-phase coefficients $I^{\mathrm{Re}}_{k,h}$ and $I^{\mathrm{Im}}_{k,h}$ for each $(T^{\ast},\omega)$ may be written compactly:
\begingroup
\begin{subequations}\label{ec:lex_compact}
\begin{align}
\underset{\bm{x}}{\mathrm{lex\,minimize}}\quad
& \tau,J_{\mathrm{SCL}}\label{ec:lex_compact_a}\\
\text{subject to}\quad
& i_{1}(t)=0 \quad\forall t\label{ec:open_phase_const}\\
& \frac{1}{n_{\mathrm{t}}}\sum_{t}T(t)=T^{\ast}\label{ec:lex_compact_b}\\
& \bm{1}_{1\times6}\:\bm{i}(t)=0\quad\forall t\label{ec:lex_compact_c}\\[-0.2ex]
	& i_{k}(t) \leq  i^{\mathrm{mx}}_{\mathrm{pk}} \quad\forall k,\ \forall t\label{ec:lex_compact_d}\\
& v^{\prime}_{r}(t) \leq  v^{\mathrm{mx}}_{\mathrm{pk}} \quad\forall r,\ \forall t\label{ec:lex_compact_e}\\
  & T(t)\leq z_{1} \quad\forall t\label{ec:lex_compact_f}\\
	& T(t)\geq z_{2} \quad\forall t\label{ec:lex_compact_g}\\
	& \tau\leq\tau^{\mathrm{mx}},\quad\text{where }
  \tau=z_{1}-z_{2}.\label{ec:lex_compact_h}
\end{align}
\end{subequations}
\endgroup

In (\ref{ec:lex_compact_a}), ``lex minimize'' means that lexicographic minimization is applied, i.e., $\tau$ is minimized first and then, among all solutions attaining the minimum $\tau$, $J_{\mathrm{SCL}}$ is minimized. $J_{\mathrm{SCL}}$ represents SCL normalized by stator resistance:
\begin{align}
\label{ec:J}
J_{\mathrm{SCL}}=\sum_{k=2}^{6} i_{k,\mathrm{rms}}^2=\frac{1}{2}\sum_{k=2}^{6}\sum_{h}\left[\left(I^{\mathrm{Re}}_{k,h}\right)^{2}+\left(I^{\mathrm{Im}}_{k,h}\right)^{2}\right].
\end{align} 
This lexicographic priority is adopted because torque smoothness is often the primary requirement for the drive \cite{Atallah2003TIA,Xiong2020TIE,Yepes2024TTE}, to mitigate vibration, acoustic noise, mechanical stress, and low-frequency speed oscillations. 

The optimization variable vector $\bm{x}$ in (\ref{ec:lex_compact_a}) contains the current Fourier coefficients of all phases except $k=1$ (phase~a is open) as $\bm{x}_{\mathrm{i}}$, plus auxiliary variables $z_{1}$ and $z_{2}$:
\begin{equation}\label{ec:x}
\bm{x}=\left[\begin{array}{ccc}
\bm{x}_{\mathrm{i}}^{\mathrm{T}} & z_{1} & z_{2}
\end{array}\right]^{\mathrm{T}}
\in\mathbb{R}^{\left(10n_{\mathrm{h}}+2\right)\times 1}
\end{equation}
where
{\setlength{\arraycolsep}{3pt}\begin{align}
\bm{x}_{\mathrm{i}}&=\left[
\begin{array}{cccccc}
\bm{x}_{2}^{\mathrm{T}} & \bm{x}_{3}^{\mathrm{T}} & \cdots & \bm{x}_{k} & \cdots & \bm{x}_{6}^{\mathrm{T}}
\end{array}
\right]^{\mathrm{T}}
\in\mathbb{R}^{10n_{\mathrm{h}}\times 1}\label{ec:xi}\\[-1mm]
\bm{x}_{k}&=\left[\begin{array}{cccccc}
\bm{x}_{k,1}^{\mathrm{T}} & \bm{x}_{k,3}^{\mathrm{T}} & \cdots & \bm{x}_{k,h}^{\mathrm{T}} & \cdots & \bm{x}_{k,H}^{\mathrm{T}}
\end{array}\right]^{\mathrm{T}}\in\mathbb{R}^{2n_{\mathrm{h}}\times 1}\label{ec:xk}\\[-1mm]
\bm{x}_{k,h}&=\left[\begin{array}{cc}I^{\mathrm{Re}}_{k,h} & I^{\mathrm{Im}}_{k,h}\end{array}\right]^{\mathrm{T}}.\label{ec:xkh}
\end{align}}%

The time variable $t$ takes in (\ref{ec:lex_compact}) values according to $t\in\left\{t_{1}, t_{2}, \ldots , t_{n_{\mathrm{t}}}\right\}$, where $t_{1}, t_{2}, \ldots , t_{n_{\mathrm{t}}}$ are discrete instants within a fundamental period, and $n_{\mathrm{t}}$ must be sufficiently large to ensure an accurate evaluation of the peak-based constraints.



The constraint (\ref{ec:open_phase_const}) sets the current of the open phase to zero in the optimization, so that this condition is taken into account when evaluating the constraints and objective functions.

Constraint (\ref{ec:lex_compact_b}) imposes equality between the average value of the torque $T(t)$ [see (\ref{ec:Tem})] and its reference $T^{\ast}$. Constraint (\ref{ec:lex_compact_c}) enforces the single-neutral zero-sum condition ($i_{0^{\sumsmallb}}=0$), where $\bm{1}_{\mu \times \nu}$ is a $\mu$$\times$$\nu$ all-ones matrix. Constraints (\ref{ec:lex_compact_d}), (\ref{ec:lex_compact_e}) and (\ref{ec:lex_compact_f})-(\ref{ec:lex_compact_h}) limit, respectively, the instantaneous healthy-phase currents $i_{k}(t)$, the instantaneous line voltages $v^{\prime}_{r}(t)$ physically constrained by the dc-link voltage, and the torque ripple $\tau$ to the respective limits $i^{\mathrm{mx}}_{\mathrm{pk}}$, $v^{\mathrm{mx}}_{\mathrm{pk}}$ and $\tau^{\mathrm{mx}}$. 



In (\ref{ec:lex_compact_d}) and (\ref{ec:lex_compact_e}), the positive-valued constraints also bound the negative peaks of the phase-current and voltage signals. This is because these signals have half-wave symmetry, assuming that only odd current harmonics $h$ are considered and the back-EMF waveforms also satisfy half-wave symmetry.

In (\ref{ec:lex_compact_e}), $v'_{r}(t)$ denotes the $r$th component of
$\bm{v}'(t)$, i.e., $v'_r(t)=[\bm{v}'(t)]_r$, and represents the $r$th
line-voltage signal to be limited by the dc-link voltage. Under phase-a
open operation, the phase voltages become unbalanced, and hence all
line voltages between healthy terminals must be checked:
\begin{equation}
\bm{v}'(t)=\bm{G}'\bm{v}(t)
=
\begin{bmatrix}
v'_1(t) & v'_2(t) & \cdots & v'_{10}(t)
\end{bmatrix}^{\mathrm{T}}
\in\mathbb{R}^{10\times1}
\end{equation}
with $\bm{G}'$ selecting, in order, the ten signals
$v'_1(t)=v_2(t)-v_3(t)$, $v'_2(t)=v_2(t)-v_4(t)$, $\ldots$,
and $v'_{10}(t)=v_5(t)-v_6(t)$:
\begin{equation}
\bm{G}^{\prime}=
\begin{bmatrix}
0 & 1 & -1 & 0 & 0 & 0\\
0 & 1 & 0 & -1 & 0 & 0\\
0 & 1 & 0 & 0 & -1 & 0\\
0 & 1 & 0 & 0 & 0 & -1\\
0 & 0 & 1 & -1 & 0 & 0\\
0 & 0 & 1 & 0 & -1 & 0\\
0 & 0 & 1 & 0 & 0 & -1\\
0 & 0 & 0 & 1 & -1 & 0\\
0 & 0 & 0 & 1 & 0 & -1\\
0 & 0 & 0 & 0 & 1 & -1
\end{bmatrix}
\in\mathbb{R}^{10\times 6}.
\label{ec:Gprime_phase_a_open}
\end{equation} 

In (\ref{ec:lex_compact_f}) and (\ref{ec:lex_compact_g}),  $z_{1}$ and $z_{2}$  are auxiliary variables representing, respectively, upper and lower torque envelopes, so that the torque ripple is $\tau=z_{1}-z_{2}$ in (\ref{ec:lex_compact_h}), as done in \cite{Yepes2026TIE}.

Since $T_{\mathrm{cog}}(t)$ is included in $T(t)$ in (\ref{ec:Tem}), cogging torque is accounted for in the optimization through the mean-torque constraint (\ref{ec:lex_compact_b}) and the torque-envelope constraints (\ref{ec:lex_compact_f})--(\ref{ec:lex_compact_h}).



To make the optimization problem (\ref{ec:lex_compact}) easier to solve, it is next decomposed into two consecutive stages, as shown in Fig.~\ref{fig:offline_flowchart}: first, a convex linear program that minimizes $\tau$; and second, a convex quadratic program that minimizes the SCL. 

\subsubsection{First-Stage Linear Program for Minimum Torque Ripple}
\label{sec:LP_stage}

For each  $(T^{\ast},\omega)$, the minimum achievable sampled torque ripple $\tau^{\min}$ is first found by solving the linear program
\begingroup
\begin{subequations}\label{ec:LP}
\begin{align}
\underset{\bm{x}}{\mathrm{minimize}}\quad & \bm{c}_{\tau}^{\mathrm{T}}\bm{x}\label{ec:LP_a}\\
\text{subject to}\quad &\bm{A}_{\mathrm{eq}}\bm{x}=\bm{b}_{\mathrm{eq}}\label{ec:LP_b}\\
&\bm{A}_{\mathrm{ineq}}\bm{x}\leq\bm{b}_{\mathrm{ineq}}\label{ec:LP_c}
\end{align}
\end{subequations}\vspace{-5pt}
\endgroup
where
\begin{equation}\label{ec:ctau}
\bm{c}_{\tau}=\left[\begin{array}{ccc}
\bm{0}_{1\times 10n_{\mathrm{h}}} & 1 & -1
\end{array}\right]^{\mathrm{T}}
\in\mathbb{R}^{(10n_{\mathrm{h}}+2)\times1}
\end{equation}
with $\bm{0}_{\mu \times \nu}$ denoting a null matrix of size $\mu$$\times$$\nu$. Let $\bm{x}_{\tau}$ denote an optimal solution of (\ref{ec:LP}). The optimum value
\begin{equation}\label{ec:taumin}
\tau^{\min}=\bm{c}_{\tau}^{\mathrm{T}}\bm{x}_{\tau}
\end{equation}
is the minimum $\tau$ attainable while satisfying the mean-torque, current-sum, peak-current and peak-voltage constraints. The operating point is discarded if $\tau^{\min}>\tau^{\mathrm{mx}}$. 

The equality constraint in (\ref{ec:LP_b}) corresponds to (\ref{ec:lex_compact_b}) and (\ref{ec:lex_compact_c}), and the inequality constraint in (\ref{ec:LP_c}) to (\ref{ec:lex_compact_d})--(\ref{ec:lex_compact_g}), both written linearly as a function of $\bm{x}$. The matrices/vectors $\bm{A}_{\mathrm{eq}}$, $\bm{A}_{\mathrm{ineq}}$, $\bm{b}_{\mathrm{eq}}$ and $\bm{b}_{\mathrm{ineq}}$ are given in the Appendix. Constraint (\ref{ec:open_phase_const}) is not explicit in (\ref{ec:LP}) because $i_{1}(t)=0$ is already embedded in the objective and other constraints. Since the objective and constraints are linear, (\ref{ec:LP}) is a convex linear program. Although MATLAB $\mathrm{fmincon}$ could be used, $\mathrm{linprog}$ better exploits the linear structure, improving robustness and efficiency while yielding a global optimum.



\subsubsection{Second-Stage Quadratic Program for Minimum SCL}
\label{sec:QP_stage}

If $\tau^{\min}\leq\tau^{\mathrm{mx}}$, the final current Fourier coefficients for the considered $(T^{\ast},\omega)$ are obtained by minimizing the SCL while constraining $\tau$ to the $\tau^{\min}$ value found in the first stage. The resulting quadratic program is
\begin{subequations}\label{ec:QP}
\begin{align}
\underset{\bm{x}}{\mathrm{minimize}} \quad & J_{\mathrm{SCL}}(\bm{x})=\frac{1}{2}\bm{x}^{\mathrm{T}}\bm{Q}\bm{x}  \label{ec:subeq_a} \\
\text{subject to} \quad &\bm{A}_{\mathrm{eq}}\bm{x}=\bm{b}_{\mathrm{eq}} \label{ec:subeq_b} \\
& \bm{A}^{\prime}_{\mathrm{ineq}}\bm{x}\leq\bm{b}^{\prime}_{\mathrm{ineq}} \label{ec:subeq_c}
\end{align}%
\end{subequations}%
where $\bm{A}_{\mathrm{eq}}$ and $\bm{b}_{\mathrm{eq}}$ are the same as in (\ref{ec:LP}), and
\begin{align}\label{ec:Aineq}
\bm{A}^{\prime}_{\mathrm{ineq}}&=\left[\begin{array}{c}
\bm{A}_{\mathrm{ineq}}\\
\bm{A}^{\tau}_{\mathrm{ineq}}
\end{array}\right]
\in\mathbb{R}^{(17n_{\mathrm{t}}+1)
\times(10n_{\mathrm{h}}+2)}\\[-1mm]
\bm{b}^{\prime}_{\mathrm{ineq}}&=\left[\begin{array}{c}
\bm{b}_{\mathrm{ineq}}\\
b^{\tau}_{\mathrm{ineq}}
\end{array}\right]
\in\mathbb{R}^{(17n_{\mathrm{t}}+1)\times1}.\label{ec:bineq}
\end{align}
$\bm{A}_{\mathrm{ineq}}$ and $\bm{b}_{\mathrm{ineq}}$ are as in (\ref{ec:LP}). $\bm{A}^{\tau}_{\mathrm{ineq}}$ and $b^{\tau}_{\mathrm{ineq}}$ impose $\tau\leq\tau^{\min}+\lambda_{1}\approx\tau^{\min}$ in (\ref{ec:subeq_c}) linearly as a function of $\bm{x}$:
\begin{align}\label{ec:Atau}
\bm{A}^{\tau}_{\mathrm{ineq}}&=\bm{c}_{\tau}^{\mathrm{T}}=\left[\begin{array}{ccc}
\bm{0}_{1 \times 10n_{\mathrm{h}}} & 1 & -1
\end{array}\right]
\in\mathbb{R}^{1\times(10n_{\mathrm{h}}+2)}\\
b^{\tau}_{\mathrm{ineq}}&=\tau^{\min}+\lambda_{1}\label{ec:btau}
\end{align}
where $\lambda_{1}>0$ is a small tolerance to improve numerical robustness while preserving the priority of $\tau^{\min}$. It can be set to an insignificant value in terms of practical torque ripple (e.g., $0.01\Nm$) but not extremely low, so that the SCL is not unnecessarily increased. In any case, $\lambda_{1}$ may be neglected compared with $\tau^{\mathrm{mx}}$. The Hessian matrix in (\ref{ec:subeq_a}) is
\begin{equation}\label{ec:Q}
\bm{Q}=\left[\begin{array}{cc}
\pmb{\mathbb{I}}_{10n_{\mathrm{h}}} & 0 \\
0 & \lambda_{2}\pmb{\mathbb{I}}_{2}
\end{array}\right]
\in\mathbb{R}^{\left(10n_{\mathrm{h}}+2\right)\times\left(10n_{\mathrm{h}}+2\right)}
\end{equation}
with $\pmb{\mathbb{I}}_{m}$ being an $m$$\times$$m$ identity matrix, and $\lambda_{2}>0$ being a small (e.g., $10^{-6}$) Tikhonov regularization weight to eensure strict convexity and hence a unique solution, and to improve numerical robustness \cite{Yepes2026TIE}. Given that $\bm{Q}$ is positive definite and the constraints are linear, (\ref{ec:QP}) is indeed a convex quadratic program, as intended, and can be solved efficiently to global optimality with conventional quadratic-program solvers, e.g., using MATLAB $\mathrm{quadprog}$. The resulting optimum $\bm{x}$, containing the optimum $I^{\mathrm{Re}}_{k,h}$ and $I^{\mathrm{Im}}_{k,h}$ for the considered operating point, is called $\bm{x}_{\mathrm{opt}}$.

Although the first-stage solution $\bm{x}_{\tau}$ may not be unique, the second-stage problem (\ref{ec:QP}) does not depend on the particular $\bm{x}_{\tau}$ obtained, but only on the optimal value $\tau^{\min}$. Thus, first-stage nonuniqueness is not a problem, since (\ref{ec:QP}) selects the minimum-SCL point within the near-minimum-$\tau$ feasible set.


\subsubsection{Description of the LUT Generation Process}
\label{sec:LUT}

The offline flowchart in Fig.~\ref{fig:offline_flowchart} generates the LUTs of $I^{\mathrm{Re}}_{k,h}$ and $I^{\mathrm{Im}}_{k,h}$ by solving the lexicographic problem over a grid of $(T^{\ast},\omega)$ points. For each $T^{\ast}$, $\omega$ is swept from zero in $\Delta\omega$ steps until infeasibility. $T^{\ast}$ is then swept from zero in $\Delta T$ steps until no solution is found even for $\omega\approx0$. The resulting coefficients are saved in the LUTs. Finally, since they are speed-invariant while the voltage constraints are inactive, the entries for $\omega<\min(\bm{\omega}^{\downarrow})$ are removed to reduce the LUTs, where $\bm{\omega}^{\downarrow}$ collects, for each $T^{\ast}$, the lowest speed at which any $v_{\mathrm{pk}}$ constraint is active.

\section{Simulation Results}
\label{sec:sims}

\begin{figure}[t!]
\centering
\includegraphics[width=0.85\columnwidth]{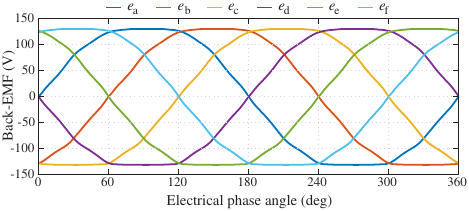}
\caption{Phase back-EMF voltage waveforms of the considered PMSM at $1100\rpm$ over one fundamental electrical cycle.}
\label{fig:backEMF_actual_PMSM}
\end{figure}

The model of the real symmetrical six-phase nonsalient 12-slot/10-pole PMSM used in \cite{Yepes2026TIE} is considered here for simulation analysis, using star connection and phase~a open. Its rated current, voltage, torque, and speed are $3.2\A$, $100\V$, $16.7\Nm$, and $1100\rpm$, respectively. In the VSD model (see Section~\ref{sec:model}), the stator inductances associated with the $\alpha\beta$, $xy$, and $0^{-}$ subspaces are $12\mH$, $11.3\mH$, and $9.4\mH$, respectively. The stator resistance is $1.4\ohm$ in every VSD axis. The phase back-EMF voltage waveforms at $1100\rpm$ over one fundamental electrical cycle are shown in Fig.~\ref{fig:backEMF_actual_PMSM}.

The limits specified for the proposed method are, as in \cite{Yepes2026TIE}, as follows: $v^{\mathrm{mx}}_{{\mathrm{pk}}}=290\V$, $i^{\mathrm{mx}}_{{\mathrm{pk}}}=4.34\A$, $\tau^{\mathrm{mx}}=2\Nm$. 

\subsection{Compensation of Cogging Torque}

Next, Ansys Maxwell finite-element analysis (FEA) evaluates cogging-torque compensation by the proposed method.

\begin{figure}[t!]
\centering
\subfloat[\label{fig:OPFa_Tref5_without_Tcog_current_torque}]{\includegraphics[width=0.9\columnwidth]{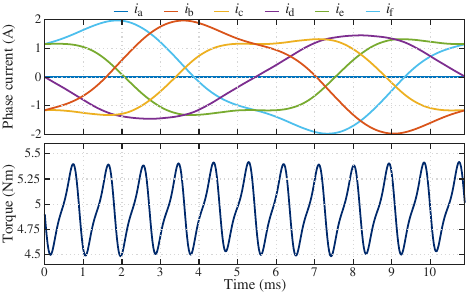}}\\
\subfloat[\label{fig:OPFa_Tref5_with_Tcog_current_torque}]{\includegraphics[width=0.9\columnwidth]{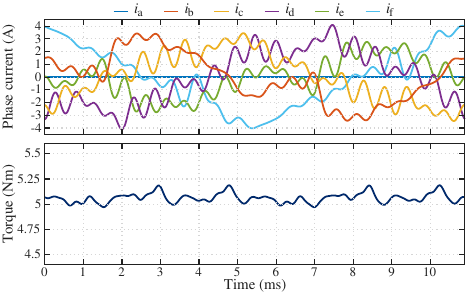}}
\caption{Evaluation of current references under phase~a open in presence of cogging torque by FEA, when the current references are obtained by the proposed method either (a)~ignoring the cogging torque [$T_{\mathrm{cog}}(t)=0$ is assumed] or (b)~including in $T_{\mathrm{cog}}(t)$ the cogging torque obtained at no load.}
\label{fig:OPFa_Tref5_Tcog}
\end{figure}

First, current references are generated with the proposed method for phase~a open ($i_{\mathrm{a}}=0$) for $T^{\ast}=5\Nm$ and $1100\rpm$ with $H=21$ and $n_{\mathrm{t}}=450$, assuming \mbox{$T_{\mathrm{cog}}(t)=0$}. Fig.~\ref{fig:OPFa_Tref5_without_Tcog_current_torque} shows that, at steady state, these currents prevent the typical open-phase low-frequency torque oscillations, ensuring fault tolerance. However, a 12th-order torque-ripple component remains, mainly due to cogging torque.


Next, Maxwell is run with zero stator currents at the same speed, and the resulting torque waveform [similar to the $T(t)$ ac part in Fig.~\ref{fig:OPFa_Tref5_without_Tcog_current_torque}] is stored. The lexicographic optimization problem is afterward solved while using this waveform as $T_{\mathrm{cog}}(t)$ to generate suitable current references. As shown in Fig.~\ref{fig:OPFa_Tref5_with_Tcog_current_torque}, the resulting currents, for the same conditions as in Fig.~\ref{fig:OPFa_Tref5_without_Tcog_current_torque}, include greater harmonic content, but remain below the specified current limit $i^{\mathrm{mx}}_{\mathrm{pk}}=4.34\A$. Provided a current controller capable of accurately tracking the harmonic references is available \cite{Kestelyn2011TIE,Mohammadpour2015TIA,Yepes2024TTE,Yepes2026TIE,Cervone2021TPEL}, these currents achieve an effective reduction in peak-to-peak $\tau$ from $0.94\Nm$ to just $0.22\Nm$ (see Fig.~\ref{fig:OPFa_Tref5_Tcog}). The remaining small $\tau$ may be due to neglected effects such as reluctance torque and flux saturation, which may be addressed in future work.

\subsection{Evaluation of Feasible Torque-Speed Range}

\begin{figure}[t!]
\centering
\includegraphics[width=0.9\columnwidth]{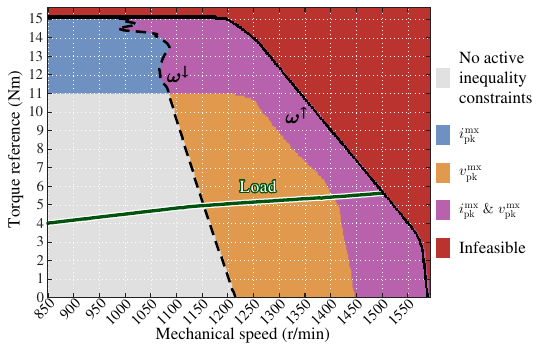}
\caption{Active inequality constraints over the feasible torque-speed regions, and infeasible region, for the proposed method under phase~a open.}
\label{fig:active_constraints_map}
\end{figure}

Using the aforesaid parameters and limits, the feasible $T$-$\omega$ range is evaluated in MATLAB for phase~a open. $T_{\mathrm{cog}}(t)=0$, $H=21$ and $n_{\mathrm{t}}=250$ are used for this and the following figures. Setting a conservative breakpoint resolution of $0.1\Nm$ and $1.2\rpm$ \cite{Yepes2026TIE}, each LUT has a size of $303$$\times$$532$. Fig.~\ref{fig:active_constraints_map} shows the feasible region and active inequality constraints: gray means no active inequality constraint, and colored areas indicate active $i_{\mathrm{pk}}$ and/or $v_{\mathrm{pk}}$ limits. The region is limited by $i^{\mathrm{mx}}_{\mathrm{pk}}$ and $v^{\mathrm{mx}}_{\mathrm{pk}}$ at high $T$ and $\omega$, respectively.

\begin{figure}[t!]
\centering
\includegraphics[width=0.95\columnwidth]{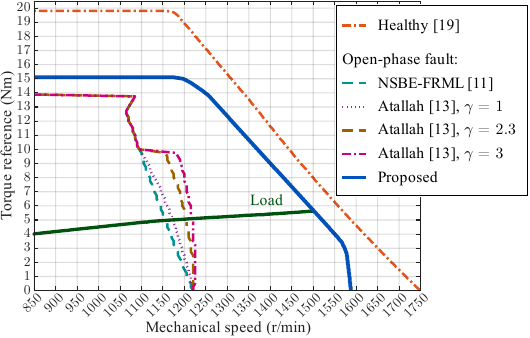}
\caption{Feasible $T$-$\omega$ limit curve of the proposed method under phase~a open, compared with the healthy case of \cite{Yepes2026TIE} and with NSBE-FRML \cite{Yepes2024TTE} and Atallah \emph{et al.} \cite{Atallah2003TIA} evaluated for phase~a open. For Atallah \emph{et al.}, three values of the flux-weakening weighting-factor multiplier $\gamma$ \cite{Sun2010TIE,Yepes2026TIE} are shown.}
\label{fig:omega_uparrow_comparison}
\end{figure}

Fig.~\ref{fig:omega_uparrow_comparison} compares the feasible $T$-$\omega$ boundary with the healthy case of \cite{Yepes2026TIE} and previous open-phase methods. NSBE-FRML \cite{Yepes2024TTE} and the method of Atallah \emph{et al.} \cite{Atallah2003TIA} are evaluated under the same phase-a open condition by setting $i_{\mathrm{a}}=0$ and enforcing the isolated-neutral current-sum constraint. For the latter \cite{Atallah2003TIA}, several $\gamma$ values are considered, with $\gamma$ scaling its flux-weakening weighting factor \cite{Sun2010TIE,Yepes2026TIE}.

Fig.~\ref{fig:omega_uparrow_comparison} confirms that including the voltage constraint and all healthy-phase Fourier coefficients substantially enlarges the postfault torque-speed region. Previous postfault methods reach only about $1220$--$1225\rpm$ at low torque, even after increasing $\gamma$ in the method of Atallah \emph{et al.}; by contrast, the proposed method remains feasible up to about $1588\rpm$ at low torque. The remaining gap with the healthy benchmark of \cite{Yepes2026TIE} ($1748\rpm$, $19.8\Nm$) is due to the phase-a open fault, which inevitably reduces the drive capability; nevertheless, the proposed method keeps this reduction moderate.

\subsection{Evaluation of Metrics Over Example Load Curve}

\begin{figure}[t!]
\centering
\includegraphics[width=0.95\columnwidth]{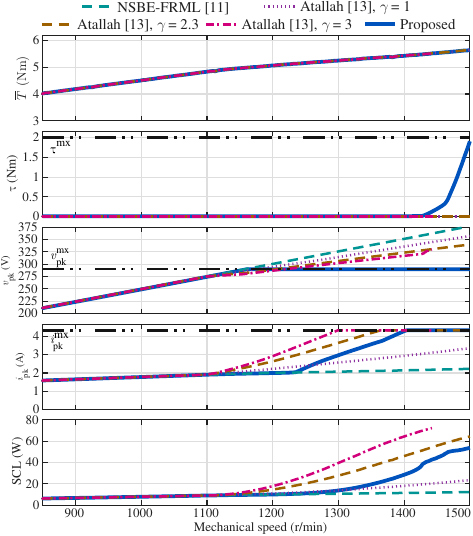}
\caption{From top to bottom: average torque, peak-to-peak torque ripple, peak line voltage, peak phase current, and SCL along the example load curve (see Figs.~\ref{fig:active_constraints_map} and~\ref{fig:omega_uparrow_comparison}) under phase~a open, comparing the proposed method with NSBE-FRML \cite{Yepes2024TTE} and Atallah \emph{et al.} \cite{Atallah2003TIA} for several values of the flux-weakening weighting-factor multiplier $\gamma$.}
\label{fig:i_pk_load_transient}
\end{figure}

Fig.~\ref{fig:i_pk_load_transient} evaluates the main metrics along the example load curve of Figs.~\ref{fig:active_constraints_map} and~\ref{fig:omega_uparrow_comparison}. Here, $\overline{T}$ is cycle-average torque, $v_{\mathrm{pk}}=\max_{\theta_{\mathrm{r}}}\|\bm{v}^{\prime}(\theta_{\mathrm{r}})\|_{\infty}$, and $i_{\mathrm{pk}}=\max_{k,\theta_{\mathrm{r}}}|i_k(\theta_{\mathrm{r}})|$, with these maxima being taken within one electrical cycle at each load-curve point. NSBE-FRML yields small SCL, but soon exceeds $v^{\mathrm{mx}}_{\mathrm{pk}}$ because voltage is not constrained. Increasing $\gamma$ in the method of Atallah \emph{et al.} reduces $v_{\mathrm{pk}}$ \cite{Sun2010TIE}, but at the cost of higher current and SCL. Moreover, $v_{\mathrm{pk}}$ still surpasses $v^{\mathrm{mx}}_{\mathrm{pk}}$ well below $1500\rpm$ even in the best case, at about $1220\rpm$. For $\gamma=3$, the line is interrupted before reaching $1500\rpm$ because the current limit prevents torque tracking at high speeds. In contrast, the proposed method tracks $\overline{T}$ while keeping $\tau$, $v_{\mathrm{pk}}$, and $i_{\mathrm{pk}}$ within their prescribed limits, up to $1500\rpm$, and with lower SCL than the method of Atallah \emph{et al.} despite satisfying the constraints.

\section{Conclusions}
\label{sec:conclusions}

This paper has presented a LUT-based current-reference generation method for a star-connected symmetrical six-phase nonsalient PMSM with phase~a open and nonsinusoidal back-EMF. Fourier coefficients of all healthy phases are optimized offline, allowing unbalanced nonsinusoidal currents, by a two-stage lexicographic procedure that first minimizes torque ripple and then SCL, under mean-torque, zero-current-sum, peak-current, peak-voltage, and torque-ripple constraints.


FEA results show that the method is able to substantially compensate cogging torque, provided its waveform is recorded in advance at zero load and included in the offline optimization for the LUTs. MATLAB simulation results show a larger phase-a-open torque-speed region than the considered previous methods: at low torque, the feasible speed limit increases from about $1220$--$1225\rpm$ to $1588\rpm$ (roughly $30\%$), whereas along the example load curve it increases from at most about $1220\rpm$ (best case) to $1500\rpm$ (roughly $23\%$), within all prescribed limits. Future work may generalize the approach to arbitrary phase numbers, fault scenarios, and winding arrangements, while addressing other nonidealities such as saturation, saliency, and thermal parameter drift.



\appendix[Matrices and Vectors of the Common Constraints]
\label{app:common_matrices}

This appendix gives the common matrices and vectors used in the two stages of the offline lexicographic optimization in Section~\ref{sec:offline}. They implement the constraints (\ref{ec:lex_compact_b})--(\ref{ec:lex_compact_g}) linearly in $\bm{x}$ [see (\ref{ec:x})] for the consecutive problems (\ref{ec:LP}) and (\ref{ec:QP}). They are derived based on (\ref{ec:open_phase_const})--(\ref{ec:lex_compact_g}) and the current-reference construction equation (\ref{ec:online}) combined with (\ref{ec:theta_k}). In the following, superscripts of $\bm{A}$ and $\bm{b}$ indicate the corresponding constraint equations of (\ref{ec:lex_compact}). First, let us define\vspace{-3pt}
\begin{equation}\label{ec:aT}
\bm{a}_{\mathrm{T}}(t)=\sum_{k=2}^{6}e^{\prime}_{k}(t)\bm{f}_{k}(t)
\in\mathbb{R}^{1\times 10n_{\mathrm{h}}}
\end{equation}\vspace{-\baselineskip}
\begin{align}\setlength{\arraycolsep}{3pt}
\bm{f}_{k}(t)&=\left[\begin{array}{ccc}
\bm{0}_{1\times 2(k-2)n_{\mathrm{h}}} & \bm{q}_{k}(t) & \bm{0}_{1\times 2(6-k)n_{\mathrm{h}}}
\end{array}\right]\in\mathbb{R}^{1\times 10n_{\mathrm{h}}}\raisetag{5pt}\\
\bm{q}_{k}(t)&=\left[\begin{array}{cccc}
\bm{q}_{k,1}(t) & \bm{q}_{k,3}(t) & \cdots & \bm{q}_{k,H}(t)
\end{array}\right]
\in\mathbb{R}^{1\times 2n_{\mathrm{h}}}\raisetag{5pt}\\
\bm{q}_{k,h}(t)&=\left[\begin{array}{cc}
c_{k,h}(t) & s_{k,h}(t)
\end{array}\right]\in\mathbb{R}^{1\times2}
\end{align}
where $c_{k,h}(t)$ and $s_{k,h}(t)$ are defined as in (\ref{ec:online}).


The common matrix $\bm{A}_{\mathrm{eq}}$ and vector $\bm{b}_{\mathrm{eq}}$ of the equality constraints of both stages, in (\ref{ec:LP_b}) and (\ref{ec:subeq_b}), are
\begin{align}\label{ec:Aeq_general}
\bm{A}_{\mathrm{eq}}&=\left[\begin{array}{c}
\bm{A}^{\eqsupref{ec:lex_compact_b}}_{\mathrm{eq}}\\
\bm{A}^{\eqsupref{ec:lex_compact_c}}_{\mathrm{eq}}
\end{array}\right]
\in\mathbb{R}^{(1+2n_{\mathrm{h}})\times(10n_{\mathrm{h}}+2)}\\
\bm{b}_{\mathrm{eq}}&=\left[\begin{array}{c}
b^{\eqsupref{ec:lex_compact_b}}_{\mathrm{eq}}\\
\bm{b}^{\eqsupref{ec:lex_compact_c}}_{\mathrm{eq}}
\end{array}\right]
\in\mathbb{R}^{(1+2n_{\mathrm{h}})\times 1}
\end{align}
where the submatrices and vector blocks are
\begin{align}
\bm{A}^{\eqsupref{ec:lex_compact_b}}_{\mathrm{eq}}&=
\frac{1}{n_{\mathrm{t}}}\left[\begin{array}{ccc}
\sum_{l}\bm{a}_{\mathrm{T}}(t_l) & 0 & 0
\end{array}\right]
\in\mathbb{R}^{1\times(10n_{\mathrm{h}}+2)}\label{ec:Aeq_T}
\end{align}
\begin{equation}\label{ec:Aeq_C}
\bm{A}^{\eqsupref{ec:lex_compact_c}}_{\mathrm{eq}}=
\left[\begin{array}{c}
\vdots\\
\bm{a}^{\mathrm{c}}_{h}\\
\bm{a}^{\mathrm{s}}_{h}\\
\vdots
\end{array}\right]
\in\mathbb{R}^{2n_{\mathrm{h}}\times(10n_{\mathrm{h}}+2)}
\end{equation}
\begin{equation}
b^{\eqsupref{ec:lex_compact_b}}_{\mathrm{eq}}=
T^{\ast}-\frac{1}{n_{\mathrm{t}}}\sum_{l}T_{\mathrm{cog}}(t_l);
\quad\quad
\bm{b}^{\eqsupref{ec:lex_compact_c}}_{\mathrm{eq}}=
\bm{0}_{2n_{\mathrm{h}}\times 1}.
\label{ec:Tast_equality}
\end{equation}
To derive $\bm{A}^{\eqsupref{ec:lex_compact_c}}_{\mathrm{eq}}$ in
(\ref{ec:Aeq_C}), note that
(\ref{ec:lex_compact_c}) is equivalent to
\begin{equation}\label{ec:sum_zero}
\sum_{k=2}^{6}i_k(t)=0.
\end{equation}
Then, (\ref{ec:online}) and (\ref{ec:theta_k}) are substituted into (\ref{ec:sum_zero}), and 
angle-difference trigonometric identities are used to decompose each $\cos (h(\theta_{\mathrm{r}}(t)-\phi_k))$ and $\sin (h(\theta_{\mathrm{r}}(t)-\phi_k))$ term into four cosine and sine functions. The coefficients multiplying the resulting $\cos (h\theta_{\mathrm{r}}(t))$ and $\sin (h\theta_{\mathrm{r}}(t))$ terms should be zero for all $t$ to satisfy (\ref{ec:sum_zero}). Thus, for each harmonic $h$, two rows $\bm{a}^{\mathrm{c}}_{h}$ and
$\bm{a}^{\mathrm{s}}_{h}$ are included in (\ref{ec:Aeq_C}), whose entries associated
with $\bm{x}_{k,h}$ [see (\ref{ec:xi})--(\ref{ec:xkh})], for $k\in\{2,\ldots,6\}$, are\vspace{-2pt}
\begin{align}%
\left[\bm{a}^{\mathrm{c}}_{h}\right]_{\bm{x}_{k,h}}&=
\left[\begin{array}{cc}
\cos(h\phi_{k}) & -\sin(h\phi_{k})
\end{array}\right]\label{ec:Aeq_C_cos}\\
\left[\bm{a}^{\mathrm{s}}_{h}\right]_{\bm{x}_{k,h}}&=
\left[\begin{array}{cc}
\sin(h\phi_{k}) & \cos(h\phi_{k})
\end{array}\right].\label{ec:Aeq_C_sin}
\end{align}
All other entries (columns)  of $\bm{a}^{\mathrm{c}}_{h}$ and $\bm{a}^{\mathrm{s}}_{h}$ are zero. Thus, $\bm{A}^{\eqsupref{ec:lex_compact_c}}_{\mathrm{eq}}\bm{x}=\bm{b}^{\eqsupref{ec:lex_compact_c}}_{\mathrm{eq}}$ enforces (\ref{ec:sum_zero}) for all $t$ without the redundant rows from sampling (\ref{ec:lex_compact_c}) at all $n_{\mathrm{t}}$ instants.

The common $\bm{A}_{\mathrm{ineq}}$ and $\bm{b}_{\mathrm{ineq}}$ in (\ref{ec:LP_c}) and (\ref{ec:subeq_c}), excluding the second-stage-only $\tau$ bound [(\ref{ec:Aineq}), (\ref{ec:bineq})], are
\begin{align}\label{ec:Aineq0_def}
\bm{A}_{\mathrm{ineq}}&=\left[\begin{array}{c} 
\bm{A}^{\eqsupref{ec:lex_compact_d}}_{\mathrm{ineq}} \\
\bm{A}^{\eqsupref{ec:lex_compact_e}}_{\mathrm{ineq}}\\
\bm{A}^{\eqsupref{ec:lex_compact_f}}_{\mathrm{ineq}} \\ 
\bm{A}^{\eqsupref{ec:lex_compact_g}}_{\mathrm{ineq}} 
\end{array}\right]
\in\mathbb{R}^{17n_{\mathrm{t}}\times(10n_{\mathrm{h}}+2)}\\[-1mm]
\bm{b}_{\mathrm{ineq}}&=\left[\begin{array}{c} 
\bm{b}^{\eqsupref{ec:lex_compact_d}}_{\mathrm{ineq}} \\
\bm{b}^{\eqsupref{ec:lex_compact_e}}_{\mathrm{ineq}}\\
\bm{b}^{\eqsupref{ec:lex_compact_f}}_{\mathrm{ineq}} \\
\bm{b}^{\eqsupref{ec:lex_compact_g}}_{\mathrm{ineq}} \label{ec:bineq0_def}
\end{array}\right]
\in\mathbb{R}^{17n_{\mathrm{t}}\times1}.
\end{align}
The inequality-constraint  submatrices of $\bm{A}_{\mathrm{ineq}}$ in (\ref{ec:Aineq0_def}) are
\begin{align}\label{ec:aineq1}\setlength{\arraycolsep}{4pt}
\bm{A}^{\eqsupref{ec:lex_compact_d}}_{\mathrm{ineq}} &= 
\left[
\begin{array}{ccc}
\bm{F}_{\mathrm{a}}(t_1) & \bm{0}_{5\times 1} & \bm{0}_{5\times 1}\\
\bm{F}_{\mathrm{a}}(t_2) & \bm{0}_{5\times 1} & \bm{0}_{5\times 1}\\
\vdots & \vdots & \vdots\\
\bm{F}_{\mathrm{a}}(t_{n_{\mathrm{t}}}) & \bm{0}_{5\times 1} & \bm{0}_{5\times 1}
\end{array}
\right]
\in\mathbb{R}^{5n_{\mathrm{t}}\times(10n_{\mathrm{h}}+2)}\raisetag{10pt}\\
\bm{A}^{\eqsupref{ec:lex_compact_e}}_{\mathrm{ineq}} &= 
\left[
\begin{array}{@{}c@{}}
\bm{G}(t_1) \\
\bm{G}(t_2) \\
\vdots \\
\bm{G}(t_{n_{\mathrm{t}}})
\end{array}
\right]
\in\mathbb{R}^{10n_{\mathrm{t}}\times(10n_{\mathrm{h}}+2)}\label{ec:Aineq7e}\\
\bm{A}^{\eqsupref{ec:lex_compact_f}}_{\mathrm{ineq}}&=
\left[\begin{array}{ccc}
\bm{a}_{\mathrm{T}}(t_1) & -1 & 0 \\
\bm{a}_{\mathrm{T}}(t_2) & -1 & 0 \\
\vdots & \vdots & \vdots \\
\bm{a}_{\mathrm{T}}(t_{n_{\mathrm{t}}}) & -1 & 0
\end{array}\right]
\in\mathbb{R}^{n_{\mathrm{t}}\times(10n_{\mathrm{h}}+2)}\\
\bm{A}^{\eqsupref{ec:lex_compact_g}}_{\mathrm{ineq}}&=
-\left[\begin{array}{ccc}
\bm{a}_{\mathrm{T}}(t_1) & 0 & -1 \\
\bm{a}_{\mathrm{T}}(t_2) & 0 & -1 \\
\vdots & \vdots & \vdots \\
\bm{a}_{\mathrm{T}}(t_{n_{\mathrm{t}}}) & 0 & -1
\end{array}\right]
\in\mathbb{R}^{n_{\mathrm{t}}\times(10n_{\mathrm{h}}+2)}
\label{ec:vpeak3}
\end{align}
where
\begin{align}
\bm{G}(t)&=\left[\begin{array}{ccc}
\bm{G}_{\mathrm{i}}(t) & \bm{0}_{10\times 1} & \bm{0}_{10\times 1}
\end{array}\right]
\in\mathbb{R}^{10\times(10n_{\mathrm{h}}+2)}\label{ec:gk}\\
\bm{G}_{\mathrm{i}}(t)&=\bm{G}^{\prime}\bm{D}^{-1}
\Big\{\bm{R}\bm{D}\bm{F}(t)+\omega\bm{L}\bm{D}\bm{F}^{\prime}(t)\Big\}
\in\mathbb{R}^{10\times 10n_{\mathrm{h}}}\label{ec:gi}\raisetag{8pt}
\end{align}\vspace{-0.8\baselineskip}
\begin{equation}\label{ec:Ft}
\bm{F}(t)= 
\left[
\begin{array}{@{}c@{}}
\bm{f}_{1}(t) \\
\bm{f}_{2}(t) \\
\vdots \\
\bm{f}_{6}(t)
\end{array}
\right]
,\  
\bm{F}^{\prime}(t)= 
\left[
\begin{array}{@{}c@{}}
\bm{f}^{\prime}_{1}(t) \\
\bm{f}^{\prime}_{2}(t) \\
\vdots \\
\bm{f}^{\prime}_{6}(t)
\end{array}
\right]
\in\mathbb{R}^{6\times 10n_{\mathrm{h}}}
\end{equation}
\begin{equation}\label{ec:Fa}
\bm{F}_{\mathrm{a}}(t)=
\left[
\begin{array}{@{}c@{}}
\bm{f}_{2}(t) \\
\bm{f}_{3}(t) \\
\bm{f}_{4}(t) \\
\bm{f}_{5}(t) \\
\bm{f}_{6}(t)
\end{array}
\right]
\in\mathbb{R}^{5\times 10n_{\mathrm{h}}}
\end{equation}\vspace{-1\baselineskip}
{\setlength{\arraycolsep}{3pt}\begin{align}
\bm{f}^{\prime}_{k}(t)&=\left[\begin{array}{ccc}
\bm{0}_{1\times 2(k-2)n_{\mathrm{h}}} & \bm{q}^{\prime}_{k}(t) & \bm{0}_{1\times 2(6-k)n_{\mathrm{h}}}
\end{array}\right]\in\mathbb{R}^{1\times 10n_{\mathrm{h}}\raisetag{5pt}}\\
\bm{f}_{1}(t)&=\bm{f}^{\prime}_{1}(t)=\bm{0}_{1\times 10n_{\mathrm{h}}}\label{ec:f0}
\end{align}}\vspace{-1.3\baselineskip}%
\begin{align}
\bm{q}^{\prime}_{k}(t)&=\left[\begin{array}{cccc}
\bm{q}^{\prime}_{k,1}(t) & \bm{q}^{\prime}_{k,3}(t) & \cdots & \bm{q}^{\prime}_{k,H}(t)
\end{array}\right]
\in\mathbb{R}^{1\times 2n_{\mathrm{h}}}\raisetag{5pt}\\
\bm{q}^{\prime}_{k,h}(t)&=\left[\begin{array}{cc}
-h s_{k,h}(t) & h c_{k,h}(t)
\end{array}\right]
\in\mathbb{R}^{1\times2}.
\end{align} Since open-phase Fourier current coefficients are excluded from the optimization vector $\bm{x}$ in (\ref{ec:x}), the phase-a rows of $\bm{F}(t)$ and $\bm{F}^{\prime}(t)$ are set to zero in (\ref{ec:Ft}), in accordance with (\ref{ec:f0}). $\bm{G}_{\mathrm{i}}(t)$ in (\ref{ec:gi}) is obtained based on PMSM model (\ref{ec:model}).

The inequality-constraint vector blocks of $\bm{b}_{\mathrm{ineq}}$ in (\ref{ec:bineq0_def}) are
\begin{align}
\bm{b}^{\eqsupref{ec:lex_compact_d}}_{\mathrm{ineq}}
&=i^{\mathrm{mx}}_{\mathrm{pk}}\bm{1}_{5n_{\mathrm{t}} \times 1}
\in\mathbb{R}^{5n_{\mathrm{t}}\times1}\label{ec:bineq1}\\
\bm{b}^{\eqsupref{ec:lex_compact_e}}_{\mathrm{ineq}}
&= v^{\mathrm{mx}}_{\mathrm{pk}}\bm{1}_{10n_{\mathrm{t}} \times 1}
-\omega_{\mathrm{m}}\bm{b}_{\mathrm{e}}
\in\mathbb{R}^{10n_{\mathrm{t}}\times1}\label{ec:b7e}\\
\bm{b}^{\eqsupref{ec:lex_compact_f}}_{\mathrm{ineq}}
&=-\left[
\begin{array}{c}
T_{\mathrm{cog}}(t_{1})\\
T_{\mathrm{cog}}(t_{2})\\
\vdots\\
T_{\mathrm{cog}}(t_{n_{\mathrm{t}}})
\end{array}\right]
\in\mathbb{R}^{n_{\mathrm{t}}\times1}\label{ec:bcog1}\\
\bm{b}^{\eqsupref{ec:lex_compact_g}}_{\mathrm{ineq}}
&=-\bm{b}^{\eqsupref{ec:lex_compact_f}}_{\mathrm{ineq}}
\in\mathbb{R}^{n_{\mathrm{t}}\times1}\label{ec:bcog2}
\end{align}
\begin{equation}
\bm{b}_{\mathrm{e}}=\left[
\begin{array}{c}
\bm{G}^{\prime}\bm{e}^{\prime}(t_{1}) \\
\bm{G}^{\prime}\bm{e}^{\prime}(t_{2}) \\
\vdots \\
\bm{G}^{\prime}\bm{e}^{\prime}(t_{n_{\mathrm{t}}})
\end{array}\right]
\in\mathbb{R}^{10n_{\mathrm{t}}\times1}.
\label{ec:be}
\end{equation}


\end{document}